\documentclass[12pt,aps,prd]{revtex4}

\usepackage{epsfig}
\usepackage{amssymb}
\usepackage{amsmath}
\usepackage{amsfonts}
\usepackage{graphics}

\newcommand{\mtrx}[2]{\left(\begin{array}{#1} #2 \end{array}\right)}
\newcommand{\I}[0]{{\rm i}}
\newcommand{\eps}{\varepsilon}
\newcommand{\Refs}{Refs.}
\newcommand{\Ref}{Ref.}
\newcommand{\Sec}{Sec.}
\newcommand{\tab}{Tab.}
\newcommand{\eq}{Eq.}
\newcommand{\fig}{Fig.}
\newcommand{\sstt}{\sin^2(2\theta_{13})}

\newcommand{\ie}{\emph{i.e.}}
\newcommand{\eg}{\emph{e.g.}}
\newcommand{\cf}{\emph{c.f.}}
\newcommand{\CP}{\emph{CP}}

\DeclareMathOperator{\diag}{diag}

\begin{document}

\title{Effects of non-standard interactions in the MINOS experiment}

\author{Mattias Blennow}
\email[]{emb@kth.se}
\author{Tommy Ohlsson}
\email[]{tommy@theophys.kth.se}
\author{Julian Skrotzki}
\email[]{skrotzki@kth.se}
\affiliation{Department of Theoretical Physics, School of Engineering Sciences, Royal Institute of Technology (KTH) -- AlbaNova University Center, Roslagstullsbacken 21, 106 91 Stockholm, Sweden}

\begin{abstract}
We investigate the effects of non-standard interactions on the determination of the neutrino oscillation parameters $\Delta m_{31}^2$, $\theta_{23}$, and $\theta_{13}$ in the MINOS experiment. We show that adding non-standard interactions to the analysis leads to an extension of the allowed parameter space to larger values of $\Delta m_{31}^2$ and smaller $\theta_{23}$, and basically removes all predictability for $\theta_{13}$. In addition, we discuss the sensitivities to the non-standard interaction parameters of the MINOS experiment alone. In particular, we examine the degeneracy between $\theta_{13}$ and the non-standard interaction parameter $\eps_{e\tau}$. We find that this degeneracy is responsible for the removal of the $\theta_{13}$ predictability and that the possible bound on $|\eps_{e\tau}|$ is competitive with direct bounds only if a more stringent external bound on $\theta_{13}$ is applied.
\end{abstract}


\maketitle

\section{Introduction}

In the years that have passed since neutrino oscillations were first observed at the Super-Kamiokande experiment in 1998 \cite{Fukuda:1998mi}, there has been a remarkable experimental development in neutrino physics. For example, from the results of solar \cite{Fukuda:2002pe,Ahmad:2002jz,Ahmad:2001an} and long-baseline reactor \cite{Eguchi:2002dm,Araki:2004mb} neutrino experiments, we now know that the solution to the solar neutrino problem is given by the large mixing angle (LMA) solution with $\Delta m_{21}^2 \simeq 8\cdot 10^{-5}$ eV$^2$ and $\theta_{12} \simeq 33.2^\circ$, and from  atmospheric \cite{Fukuda:1998mi} and accelerator \cite{Ahn:2002up,Aliu:2004sq,Michael:2006rx} neutrino experiments, we know that $|\Delta m_{31}^2| \simeq 2.5\cdot 10^{-3}$ eV$^2$ and that $\theta_{23}$ is close to maximal (\ie, $\theta_{23} = \pi/4$). In addition, analyses of the $L/E$ binned Super-Kamiokande \cite{Ashie:2004mr} and KamLAND \cite{Araki:2004mb} data even show the oscillatory behavior of the neutrino flavor conversion probability.

The fact that neutrino oscillations occur implies that neutrinos have non-zero masses, which requires physics beyond the standard model of particle physics (SM). Thus, neutrino physics seems to be a viable window to explore physics beyond the SM. A feature of many extensions of the SM is the existence of non-standard interactions (NSI) (see, \eg, \Ref~\cite{Valle:2003uv} for a recent review) between neutrinos and other fermions, including the first generation fermions which make up most of the matter that we experience in everyday life. In particular, effective four-fermion operators arising from such NSI will inevitably affect the dispersion relations for neutrinos propagating in matter through coherent forward scattering similar to that of the Mikheyev--Smirnov--Wolfenstein (MSW) effect \cite{Wolfenstein:1977ue,Mikheev:1986gs,Mikheev:1986wj}, which is usually considered in neutrino oscillation analyses and responsible for the conversion of solar $\nu_e$ into $\nu_\mu$ and $\nu_\tau$. With new generations of neutrino oscillation experiments in the planning stages, we expect to probe the yet unknown parts of the parameter space for neutrino oscillations and to decrease the experimental uncertainty in the parts where we have only pinpointed certain regions. In such precision experiments, it may happen that even small contributions of NSI to the matter effects can play a role in distorting the measurements of the standard neutrino oscillation parameters or, more excitingly, that NSI can even be observed through the very same effects \cite{Huber:2002bi,Blennow:2005qj}.

The Main Injector Neutrino Oscillation Search (MINOS) \cite{Michael:2006rx} is an accelerator based neutrino oscillation experiment with a baseline of 750 km reaching from Fermilab, Illinois to the Soudan mine, Minnesota in the United States. It is an experiment designed to measure the neutrino oscillation parameters $\Delta m_{31}^2$ and $\theta_{23}$, but it may also improve the bound on the leptonic mixing angle $\theta_{13}$. In this paper, we discuss the implications of including NSI in the analysis of the MINOS data. We focus on how the introduction of NSI affects the experimental bounds on the standard neutrino oscillation parameters, but also discuss what bounds MINOS itself could put on the parameters $\eps_{\alpha\beta}$, which describe the NSI on a phenomenological level.

Non-standard interactions in the MINOS experiment have been previously studied by Friedland and Lunardini in \Ref~\cite{Friedland:2006pi}. While they focus on constraints which are put by the combination of MINOS and atmospheric neutrino oscillation experiments, we focus on the constraints that can be inferred from the MINOS experiment alone. In particular, we consider the $\nu_\mu \rightarrow \nu_e$ appearance channel and its implications for the leptonic mixing angle $\theta_{13}$ and the effective NSI parameter $\eps_{e\tau}$ in detail. Also in \Ref~\cite{Kitazawa:2006iq}, the effects of NSI on the $\nu_e$ appearance channel at MINOS were studied, focusing on the oscillation probability $P_{\mu e}$. One of the conclusions of \Ref~\cite{Kitazawa:2006iq} was that, in the most optimistic case, the oscillation probability will be so large that it cannot be described by the standard neutrino oscillation scenario alone, and thus, implying the existence of NSI. Our numerical simulations will show that $|\eps_{e\tau}| \simeq 2.5$ (which is above the current experimental bound) would be needed to establish NSI unless further external constraints can be put on $\sstt$.

The paper is organized as follows. In \Sec~\ref{sec:NSIoscillations}, we introduce the framework of neutrino oscillations including the effects of NSI. Section \ref{sec:analytic} deals with analytic considerations for the neutrino oscillation channels relevant to the MINOS experiment, while \Sec~\ref{sec:numeric} presents the results of our numerical treatment using the GLoBES software \cite{Huber:2004ka,Huber:2007ji}. Finally, in \Sec~\ref{sec:summary}, we summarize our results and give our conclusions.

\section{Neutrino oscillations and NSI}
\label{sec:NSIoscillations}

In this paper, we will use the standard three-flavor neutrino oscillation framework with an effective vacuum Hamiltonian given by
\begin{equation}
H_0 = \frac{1}{2E} U \diag(0,\Delta m_{21}^2,\Delta m_{31}^2) U^\dagger
\end{equation}
in flavor basis. Here $E$ is the neutrino energy, $\Delta m_{ij}^2 \equiv m_i^2-m_j^2$ are the neutrino mass squared differences, $U$ is the leptonic mixing matrix \cite{Yao:2006px}
\begin{equation}
U \equiv \mtrx{ccc}{c_{13}c_{12} & c_{13}s_{12} & s_{13} e^{-\I\delta} \\
-s_{12}c_{23} - c_{12}s_{23}s_{13}e^{\I\delta} & c_{12}c_{23} - s_{12}s_{23}s_{13}e^{\I\delta} & s_{23}c_{13} \\
s_{12}s_{23}-c_{12}c_{23}s_{13}e^{\I\delta} & -c_{12}s_{23} -s_{12}c_{23}s_{13}e^{\I\delta} & c_{23} c_{13}},
\end{equation}
$c_{ij} \equiv \cos(\theta_{ij})$, $s_{ij} \equiv \sin(\theta_{ij})$, $\theta_{ij}$ are the leptonic mixing angles, and $\delta$ is the \CP-violating Dirac phase. In addition, the standard matter effect on neutrino oscillations is implemented through the effective contribution \cite{Wolfenstein:1977ue,Mikheev:1986gs,Mikheev:1986wj}
\begin{equation}
H_{\rm MSW} = \diag(\sqrt{2}G_F N_e,0,0) \equiv V \diag(1,0,0)
\end{equation}
to the vacuum Hamiltonian, where $G_F$ is the Fermi constant and $N_e$ is the electron number density.

We are interested in examining the effects of introducing NSI between neutrinos and other fermions that reduce to effective four-fermion interactions. These NSI can be described by a Lagrangian density of the form
\begin{equation}
\mathcal L_{\rm NSI} = - \frac{G_F}{\sqrt 2} \sum_{\stackrel{f=u,d,e}{a=\pm 1}} \eps_{\alpha\beta}^{fa}[\overline{\nu_\alpha}\gamma^\mu (1-\gamma_5)\nu_\beta][\overline f \gamma_\mu (1+a\gamma_5) f],
\end{equation}
where the $\eps_{\alpha\beta}^{fa}$ give the strength of the NSI. In analogy with the MSW effect, terms of this type will give an effective contribution to the neutrino oscillation Hamiltonian, which will be of the form
\begin{equation}
H_{\rm NSI} = V \mtrx{ccc}{\eps_{ee} & \eps_{e\mu} & \eps_{e\tau} \\
\eps_{e\mu}^* & \eps_{\mu\mu} & \eps_{\mu\tau} \\
\eps_{e\tau}^* & \eps_{\mu\tau}^* & \eps_{\tau\tau}},
\end{equation}
where
$$
\eps_{\alpha\beta} = \sum_{f,a} \eps^{fa}_{\alpha\beta} \frac{N_f}{N_e},
$$
$N_f$ is the number density of fermions of type $f$, and we have assumed an unpolarized medium. For bounds on the parameters $\eps_{\alpha\beta}^{fa}$, see \Refs~\cite{Davidson:2003ha,Barranco:2005ps}. Generally, the NSI involving $\nu_\mu$ are quite well constrained, while the bounds on the other NSI (\ie, $\eps_{ee}$, $\eps_{e\tau}$, and $\eps_{\tau\tau}$) are of order unity. Thus, we will focus on NSI which do not involve $\nu_\mu$ interactions. In the remainder of this paper, we will work with the effective parameters $\eps_{\alpha\beta}$, assuming them to be constant, which is a good approximation as long as the matter composition does not change significantly along the neutrino baseline. The full Hamiltonian is then given by
\begin{equation}
H = H_0 + H_{\rm MSW} + H_{\rm NSI}.
\end{equation}
Effects of this type have been previously studied in \Refs~\cite{Wolfenstein:1977ue,Bergmann:1998ft,Bergmann:1999pk,Bergmann:2000gp,Gonzalez-Garcia:2001mp,Guzzo:2001mi,Fornengo:2001pm,Huber:2001zw,Huber:2001de,Ota:2001pw,Huber:2002bi,Ota:2002na,Fogli:2002xj,Bekman:2002zk,Campanelli:2002cc,Friedland:2004pp,Gonzalez-Garcia:2004wg,Miranda:2004nb,Friedland:2004ah,Friedland:2005vy,Blennow:2005qj}.

\section{Analytic considerations}
\label{sec:analytic}

In this section, we present some analytic considerations which are valid mainly for weak NSI. These will prove useful in understanding the numeric results in the next section.

\begin{figure}
\begin{center}
\includegraphics[width=12cm]{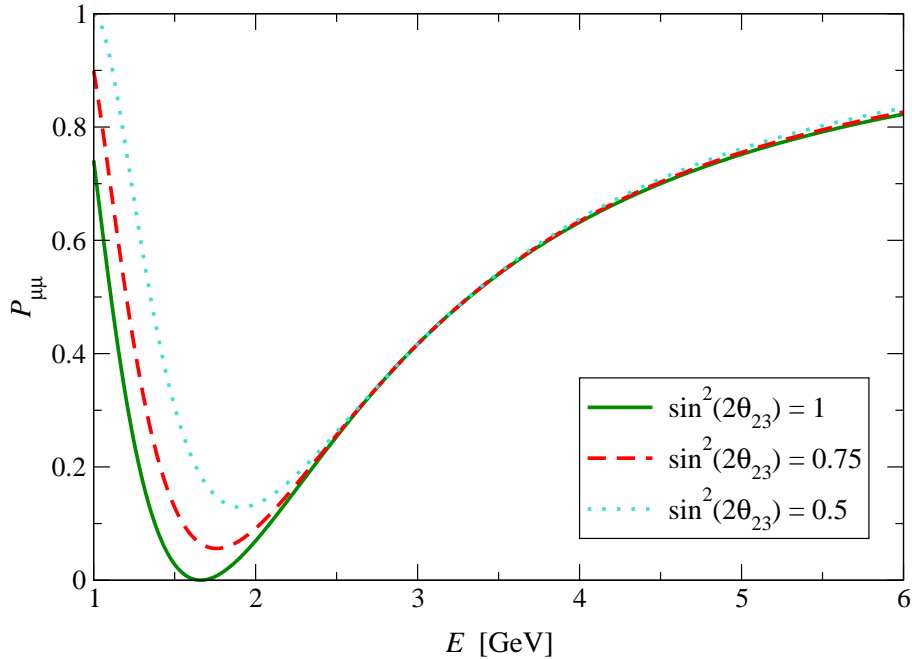}
\caption{The analytic result for the neutrino oscillation probability $P_{\mu\mu}$ as a function of the neutrino energy $E$ for different values of $\sin^2(2\theta_{23})$ and $\Delta m^2_{31}$ on the NSI degeneracy involving $\eps_{\tau\tau}$ (only the value of $\sin^2(2\theta_{23})$ is displayed in the figure) in the two-flavor scenario. The NSI parameter $\eps_{\tau\tau}$ has been chosen in order for the effective parameters to coincide at $E = 3$~GeV. \label{fig:analyticdisappearence}}
\end{center}
\end{figure}

The main objective of the MINOS experiment is to measure the neutrino oscillation parameters $\Delta m_{31}^2$ and $\theta_{23}$. The neutrino oscillation channel used is the $\nu_\mu$ disappearance channel, which is sensitive to the $\nu_\mu$ survival probability $P_{\mu\mu}$. The leading terms in the expression for $P_{\mu\mu}$ are
\begin{equation}
\label{eq:Pmm}
P_{\mu\mu} \simeq 1 - \sin^2(2\theta_{23})\sin^2\left(\frac{\Delta m_{31}^2}{4E}L\right),
\end{equation}
where three-flavor effects due to $\Delta m_{21}^2$ and $\theta_{13}$ have been neglected. Equation (\ref{eq:Pmm}) can be easily derived using the effective two-flavor Hamiltonian of the $\nu_\mu$--$\nu_\tau$ sector, \ie,
\begin{equation}
H^{\rm 2f}_0 = \frac{\Delta m_{31}^2}{4E} \mtrx{cc}{-\cos(2\theta_{23}) & \sin(2\theta_{23}) \\
\sin(2\theta_{23}) & \cos(2\theta_{23})}.
\end{equation}
For the base-line and matter potential relevant to the MINOS experiment, the off-diagonal NSI parameters will not be sufficient to introduce large transitions to $\nu_e$, and therefore, the NSI can be effectively discussed in the same two-flavor framework, where their contribution to the effective neutrino oscillation Hamiltonian is 
\begin{equation}
H^{\rm 2f}_{\rm NSI} = V\eps_{\tau\tau} \mtrx{cc}{0 & 0 \\ 0 & 1}.
\end{equation}
The stringent way of treating this situation is to introduce the parameter $\eps_{e\tau}$ as a perturbation. It is then easy to show that the two-flavor approximation holds for $|\eps_{e\tau}|^2 V^2 L^2 \ll 1$, or $|\eps_{e\tau}|^2 \ll 5.8$ in the case of the MINOS experiment.
We note that neutrino oscillations with the effective Hamiltonian $H^{\rm 2f} = H_0^{\rm 2f} + H_{\rm NSI}^{\rm 2f}$ are equivalent to standard two-flavor neutrino oscillations in matter with the substitutions $\Delta m^2 \rightarrow \Delta m_{31}^2$, $\theta \rightarrow \theta_{23}$, and $V \rightarrow -\eps_{\tau\tau} V$, for which we know that the effective neutrino oscillation parameters are given by \cite{Wolfenstein:1977ue,Mikheev:1986gs,Mikheev:1986wj}
\begin{equation}
\Delta \tilde m^2 = \Delta m_{31}^2 \xi, \quad \sin^2(2\tilde\theta) = \frac{\sin^2(2\theta_{23})}{\xi^2},
\end{equation}
where
\begin{equation}
	\xi = \sqrt{\left[\frac{2EV}{\Delta m_{31}^2} \eps_{\tau\tau} + \cos(2\theta_{23})\right]^2 + \sin^2(2\theta_{23})}.
\end{equation}
Thus, for a fixed energy $E$, it is always possible to choose $\eps_{\tau\tau}$ in such a way that $\xi = \sin(2\theta_{23})$, leading to $\Delta \tilde m^2 = \sin(2\theta_{23}) \Delta m_{31}^2$ and $\sin^2(2\tilde\theta) = 1$. Therefore, we expect, when including the effects of NSI, a degeneracy between the standard oscillation parameters $\Delta m_{31}^2$ and $\sin^2(2\theta_{23})$ and the NSI parameter $\eps_{\tau\tau}$, \ie, if we measure $\Delta\tilde m^2 = \Delta m_0^2$ and $\sin^2(2\tilde\theta) = 1$, then this could just as well be produced from a smaller mixing angle and larger mass squared difference by the effects of the NSI. The fact that different $\eps_{\tau\tau}$ will be needed in order to reproduce this effect at different energies implies that this degeneracy can be somewhat resolved by studying the neutrino oscillation probability at different energies (as in the case of an actual neutrino oscillation experiment measuring the neutrino energy, \eg, the MINOS experiment). However, if the energy range is not broad enough, then the degeneracy will still manifest itself in the form of an extension of the sensitivity contours when including NSI into the analysis. In \fig~\ref{fig:analyticdisappearence}, we show the neutrino survival probability $P_{\mu\mu}$ as a function of energy $E$ for different choices of $\sin^2(2\theta_{23})$ and $\Delta m_{31}^2$ which are on the NSI degeneracy.
Note that this figure is only provided for illustrative purposes in order to show the degeneracy and that the values of $\eps_{\tau\tau}$ needed for $\sin^2(2\theta_{23}) = 0.5$ and 0.75 are relatively large (about $-7$ for $\sin^2(2\theta_{23}) = 0.5$). However, the degeneracy does not need to be exact in order to extend the sensitivity countours and also smaller values of $\eps_{\tau\tau}$ will be enough for this purpose.

In \Ref~\cite{Blennow:2005qj}, it was shown that, including the first-order correction in $\eps_{\alpha\beta}$, the effective three-flavor mixing matrix element $\tilde U_{e3}$ is given by (to zeroth order in the ratio $\alpha = \Delta m_{21}^2/\Delta m_{31}^2$)
\begin{equation}
\label{eq:ue3approx}
\tilde U_{e3} = U_{e3} + \eps_{e\tau} \frac{2EV}{\Delta m_{31}^2} c_{23},
\end{equation}
which is valid as long as the individual contributions remain small (from the CHOOZ bound \cite{Apollonio:1999ae} $U_{e3}$ is known to be small and the absolute value of the NSI contribution is of the order of $0.2$ for $|\eps_{e\tau}| = 1$).
As the oscillation probability $P_{\mu e}$ is expected to be sensitive to the effective mixing angle $\tilde \theta_{13}$, this will lead to an additional degeneracy between the parameters $\theta_{13}$ and $\eps_{e\tau}$. The effects of the other NSI parameters are suppressed by $\alpha$ or $s_{13}$.

\section{Numeric simulations}
\label{sec:numeric}

For our numeric simulations, we used the GLoBES software \cite{Huber:2004ka,Huber:2007ji} which was extended in order to accommodate the inclusion of NSI. The Abstract Experiment Definition Language (AEDL) files used to describe the MINOS experiment were modified versions of the MINOS AEDL files provided in the GLoBES distribution and they were based on \Refs~\cite{Huber:2004ug,Ables:1995wq,NUMIL714}. These AEDL files correspond to a MINOS running time of five years with $3.7\cdot 10^{20}$ protons on target per year. The neutral- and charged-current cross-sections were taken from \Refs~\cite{Messier:1999kj,Paschos:2001np} as provided by the GLoBES distribution.

The disappearance and appearance channels were simulated in a neutrino energy interval of 1-6~GeV, since the majority of the neutrinos in the NuMI beam are in this range. For the simulations, the neutrino energy interval was binned into 30 equal bins. The matter density was assumed to be constant with a value corresponding to the matter density of the Earth's crust, \ie, $V={1}/{1900}$~km$^{-1}$. The simulated neutrino oscillation parameters are shown in \tab~\ref{tab:real_values}. The choices for $\Delta m^2_{31}$ and $\sin^2(2\theta_{23})$ are inspired by the preliminary MINOS results \cite{Michael:2006rx}, which are almost equivalent to the K2K results \cite{Aliu:2004sq}, and the simulated values of all NSI parameters are zero, in order to possibly obtain useful sensitivities for NSI detection. In all simulations, we have used the full numeric three-flavor framework and the parameters not presented in the figures (including NSI parameters such as the phase of $\eps_{e\tau}$) have been marginalized over unless stated otherwise. It should also be noted that normal mass hierarchy, \ie, $\Delta m^2_{31}>0$, was assumed for the simulations shown. The results for inverted mass hierarchy are similar and no distinction can be made between the hierarchies. Furthermore, $\sin^2(2\theta_{12})$ and $\Delta m^2_{21}$ were kept fixed at the values given in \tab~\ref{tab:real_values} for the simulations, since MINOS is not sensitive to these parameters. In addition, the \CP-violating phase $\delta$ was kept fixed for the simulations of the parameters governing the disappearance channel, since its effect in this channel is small. However, for the simulations of the parameters governing the appearance channel (\ie, $\theta_{13}$ and $\eps_{e\tau}$), we marginalize over $\delta$, since it is important to include the effects of a possible relative phase between $U_{e3}$ and $\eps_{e\tau}$. The explicit choice of $\pi/2$ for the simulated value of $\delta$ does not affect the results of our simulations significantly. In all figures, we show the combined results of the disappearance and appearance channels. Furthermore, when the standard neutrino oscillation parameters are marginalized, then we assume 20~\% external error (1$\sigma$) for $\Delta m_{31}^2$ and $\theta_{23}$ and an external error of $0.06\pi$ for $\theta_{13}$ (corresponling roughly to the CHOOZ bound).
\begin{table}
\begin{center}
\begin{tabular}{|r@{${} = {}$}l|r@{${} = {}$}l|}
\hline
$\sin^2(2\theta_{12})$ & $0.8$ & $\Delta m^2_{21}$ & $7\cdot 10^{-5}$~eV$^2$ \\
\hline
$\sin^2(2\theta_{13})$ & $0.07$ & $\Delta m^2_{31}$ & $2.74\cdot 10^{-3}$~eV$^2$ \\
\hline
$\sin^2(2\theta_{23})$ & $1$ & $\delta$ & $\frac{\pi}{2}$ \\
\hline
\end{tabular}
\end{center}
\caption{The neutrino oscillation parameters used in the simulations.}
\label{tab:real_values}
\end{table}
For the NSI parameters, we assume external errors in accordance with direct bounds \cite{Davidson:2003ha,Barranco:2005ps} and with the results of high-energy atmospheric neutrino oscillations \cite{Friedland:2004ah,Friedland:2005vy}.
The high-energy sample of atmospheric $\nu_\mu$ events indicate that muon neutrinos oscillate also at higher energies. If the second eigenvalue $\lambda_2$ of the matter interaction part of the full Hamiltonian (including both NSI and the standard matter effect) is too large (such that $|\lambda_2| \gg \Delta m_{31}^2/(2E)$), then matter effects will entirely dominate the neutrino flavor propagation. Since we have assumed $\eps_{\mu\alpha} = \eps_{\beta\mu} = 0$, this would mean that muon neutrinos would be fully decoupled in contrast to experiments. The resulting constraint in the NSI parameter space has the shape of a parabola in the $\eps_{\tau\tau}$--$|\eps_{e\tau}|$-plane as $\lambda_2 = 0$ corresponds to $\eps_{\tau\tau} = |\eps_{e\tau}|^2/(1+\eps_{ee})$ \cite{Friedland:2004ah,Friedland:2005vy}. However, the study of high-energy events of different flavors (see, \eg, \Ref~\cite{Abe:2006fu}) may provide further information such as the composition of the state which muon neutrinos oscillate into, which in turn will be related to the NSI parameters.
The atmospheric constraints were implemented by setting a prior of $|\lambda_2| < 0.2 V$ for the second eigenvalue. The results are not particularly sensitive to the specific prior chosen.

\subsection{Degeneracy of $\boldsymbol{\eps_{\tau\tau}}$}

\begin{figure}
\begin{center}
\includegraphics[width=11cm,clip=false]{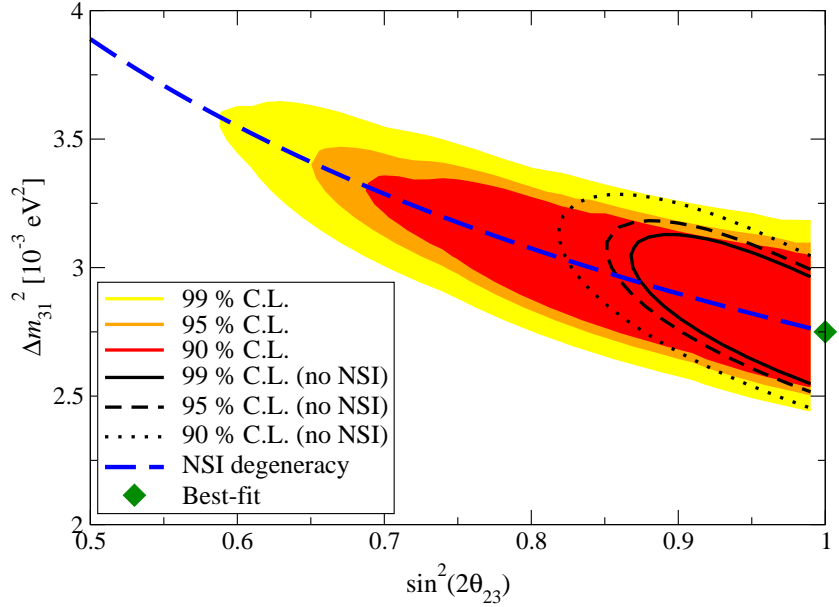}
\caption{\label{fig:s23-dm2}The sensitivity limits in the $\sin^2(2\theta_{23})$--$\Delta m_{31}^2$ plane (2 d.o.f.) for the combined appearance and disappearance channels. The colored regions correspond to the sensitivities when including NSI, while the black curves correspond to the sensitivities under the assumption that NSI are negligible. The dashed blue curve marks the NSI degeneracy involving $\eps_{\tau\tau}$ where $\Delta m^2_{31} \sin(2\theta_{23}) =  2.74\cdot 10^{-3}$ eV$^2$. The best-fit point corresponds to the parameter values used in the simulation.}
\end{center}
\end{figure}

Figure \ref{fig:s23-dm2} shows the predicted sensitivity limits of the MINOS experiment (according to the experimental setup given above) in the $\sin^2(2\theta_{23})$--$\Delta m_{31}^2$ plane with and without the inclusion of NSI. 
From this figure, we can clearly observe the extension of the sensitivity contours according to the discussion in the previous section. The reason why the contours do not extend to $\sin^2(2\theta_{23}) = 0$ is based on the fact that the MINOS experiment is not using a single neutrino energy, but rather has a continuous energy spectrum. For a fixed $\eps_{\tau\tau}$, $\xi = \sin(2\theta_{23})$ will only be fulfilled for one specific energy and the effective neutrino oscillation parameters will become energy dependent. Although $\xi = \sin(2\theta_{23})$ may still be approximately fulfilled in some finite energy range, for lower $\sin^2(2\theta_{23})$, the energy dependence will become strong enough for the MINOS experiment to detect it and this is where the sensitivity contours in \fig~\ref{fig:s23-dm2} are cut off (\cf,~\fig~\ref{fig:analyticdisappearence}). In addition, improved external bounds on the NSI parameter $\eps_{\tau\tau}$ could lead to a cutoff for the extended sensitivity contours.

\subsection{Degeneracy of $\boldsymbol{\eps_{e\tau}}$}

\begin{figure}
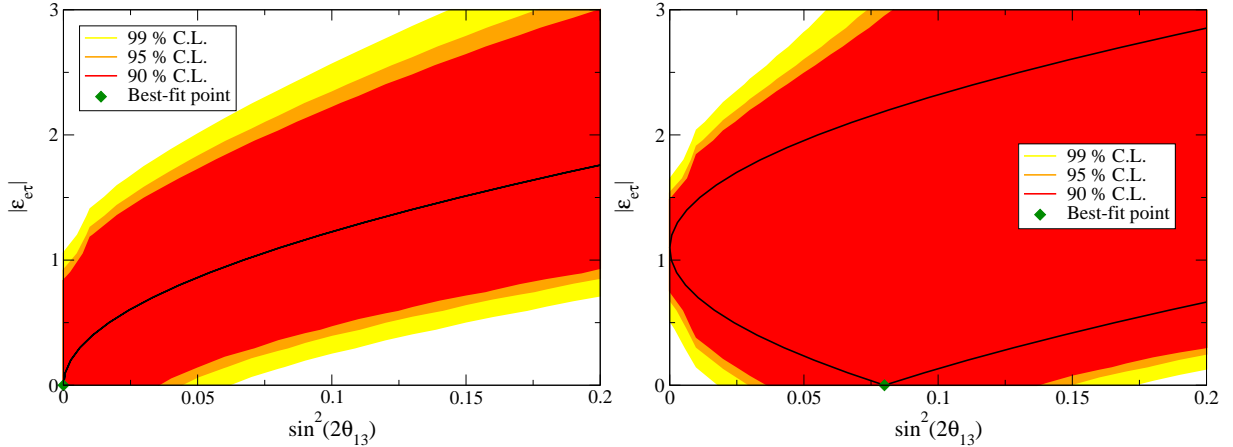

\begin{center}
\includegraphics[width=0.49\textwidth,clip=true]{s130-NSI.eps}\includegraphics[width=0.49\textwidth,clip=true]{s13-008.eps}
\caption{\label{fig:s13-et}The sensitivity limits in the $\sstt$--$|\eps_{e\tau}|$ plane (1 d.o.f.) for the combined appearance and disappearance channels. The left panel corresponds to a simulated $\sin^2(2\theta_{13}) = 0$ and the right panel to $\sin^2(2\theta_{13}) = 0.08$. The black curves correspond to $|\tilde U_{e3}|^2$ [\cf, \eq~(\ref{eq:ue3approx})] equal to the simulated value of $s_{13}^2$ for $E = 2.3$~GeV. In this figure, $\eps_{ee} = 0$ and $\eps_{\tau\tau}$ is chosen along the parabola allowed by atmospheric neutrino experiments \cite{Friedland:2004ah,Friedland:2005vy}.}
\end{center}
\end{figure}

In \fig~\ref{fig:s13-et}, we show the predicted sensitivity in the $\sstt$--$|\eps_{e\tau}|$ plane for one degree of freedom (signifying that we consider the sensitivities for the two parameters separately). This figure has been constructed assuming NSI parameters along the parabola allowed by atmospheric neutrino experiments \cite{Friedland:2004ah,Friedland:2005vy} and $\eps_{ee} = 0$ (allowing for general values of $\eps_{ee}$ slightly extend the contours). The results are shown for both $\sstt = 0$ and $\sstt = 0.08$.
As can be seen from the left panel of this figure, the MINOS experiment is sensitive to $\sstt$ which is a factor of two below the CHOOZ bound if we do not take NSI into account. However, if we include the effects of NSI, then the bound put on $\sstt$ will depend directly on the external bound on $|\eps_{e\tau}|$. Already for a bound of $|\eps_{e\tau}| \lesssim 0.5$, the bound that MINOS is able to put on $\sstt$ has deteriorated to the CHOOZ bound and is quickly getting worse for less stringent limits on $|\eps_{e\tau}|$. We can also clearly observe the $\sstt$--$|\eps_{e\tau}|$ degeneracy discussed in the previous section. The sensitivity contours of \fig~\ref{fig:s13-et} contain the degeneracy curves corresponding to $\tilde U_{e3}$ equal to the simulated values of $s_{13}^2$ for neutrino energies in the MINOS energy range. For possible bounds on $|\eps_{e\tau}|$, we need to consider some external bound on $\sstt$. With the current CHOOZ bound, the bound that could be put by the MINOS experiment is $|\eps_{e\tau}| \lesssim 2.5$, which is not competitive with the current direct bounds on the specific NSI. If the $\sstt$ bound is improved by an order of magnitude (\eg,~by future reactor experiments \cite{Ardellier:2004ui,Bolton:2005yd,Anjos:2005pg,Aoki:2006bk,Guo:2007ug}), then the MINOS bound on $|\eps_{e\tau}|$ could be improved to $|\eps_{e\tau}| \lesssim 1$, which is still of the same order of magnitude as the direct NSI bounds. Thus, it seems that in order to put constraints on NSI from neutrino oscillation experiments, we would need an experiment with better sensitivity than MINOS.

In the right panel of \fig~\ref{fig:s13-et}, we show the results for $\sstt = 0.08$ ($s_{13}^2 = |U_{e3}|^2 \simeq 0.02$). Again, we can observe the degeneracy in the $\sstt$--$|\eps_{e\tau}|$ plane along the direction of $|\tilde U_{e3}|^2 \simeq 0.02$. In this case, MINOS will be able to tell us that $\tilde U_{e3}$ is non-zero, but not whether this is the effect of non-zero $\theta_{13}$ or non-zero $\eps_{e\tau}$ (or a combination). Because of this degeneracy, the result is very similar if considering non-zero $\eps_{e\tau}$ while $\theta_{13} = 0$. When considering $\theta_{13}$ and $\eps_{e\tau}$ which are simultaneously non-zero, the resulting sensitivity contours depend on the magnitude of the simulated $\tilde U_{e3}$ [\eg, if the two terms in \eq~(\ref{eq:ue3approx}) cancel, then we obtain sensitivity contours similar to the left panel].

\subsection{Prospects of detecting NSI at MINOS}

\begin{figure}
\begin{center}
\includegraphics[height=0.8\textwidth,angle=270]{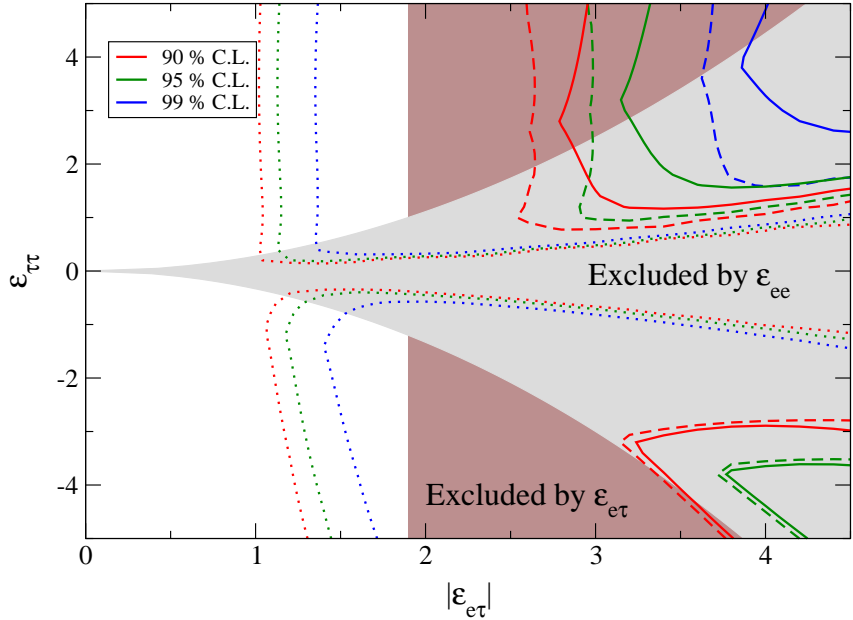}
\caption{The prospects for detecting NSI at the MINOS experiment. The solid curves correspond to our default setup with five years of running time and external bounds on $\Delta m_{31}^2$, $\theta_{23}$, and $\theta_{13}$ as defined in the beginning of this section. The dashed curves correspond to an increased running time of 15 years, whereas the dotted curves correspond to five years of running time, but with external bounds which have been improved by a factor of four. The NSI parameter $\eps_{ee}$ has been chosen along the parabola allowed by atmospheric neutrino experiments \cite{Friedland:2004ah,Friedland:2005vy}.}
\label{fig:NSIprospects}
\end{center}
\end{figure}

In \fig~\ref{fig:NSIprospects}, we show the prospects of detecting NSI at the MINOS experiment, \ie, we show the regions of the NSI parameter space where the sensitivity contours do not contain the standard oscillation scenario ($\eps_{\alpha\beta} = 0$).
As can be seen from this figure, the current situation is such that MINOS will not be sensitive to any of the NSI parameter values which are not already excluded (the excluded regions are $\eps_{ee} > 2.6$, $\eps_{ee} < -4$, and $|\eps_{e\tau}| > 1.9$ in accordance with \Ref~\cite{Davidson:2003ha}). Increasing the MINOS running time to 15 years does not improve significantly upon this result. However, if the external bounds on the standard neutrino oscillation parameters are improved by a factor of four, then MINOS will be able to detect $|\eps_{e\tau}|$ of about one at a confidence level of 90~\%. This is due to the improved bound on $\theta_{13}$ leading to a breaking of the $\sstt$--$|\eps_{e\tau}|$ degeneracy. The improvement in the external bound on $\theta_{13}$ used in the figure corresponds to an upper limit of $\sstt \simeq 10^{-2}$, which is slightly below the Double Chooz sensitivity limit \cite{Ardellier:2006mn} (note that the reactor neutrino experiments are not sensitive to NSI, since they operate at very low energies).
The asymmetry of the figure with respect to $\eps_{\tau\tau} = 0$ is a result of how $\eps_{\tau\tau}$ affects the effective $\tilde U_{e3}$. Since we are not in the perturbative regime, we treat the case with $\theta_{13} = 0$ exactly in $\eps_{\tau\tau}$ and perturbatively in $\eps_{e\tau}$. The result of this is an equation similar to \eq~(\ref{eq:ue3approx}) but where $\theta_{23}$ is an effective quantity which depends on $\eps_{\tau\tau}$. The main influence on the NSI sensitivity is given by the appearance channel, with no or only a small contribution from the disappearance channel.

\section{Summary and conclusions}
\label{sec:summary}

With the advent of new precision measurements of the neutrino oscillation parameters, it is important that we understand the phenomenology of the physics that could affect these measurements and give rise to erroneous interpretations if not taken properly into account. In addition, putting constraints on such physics from neutrino oscillation experiments alone is also an intriguing idea. In this paper, we have studied the influence of including NSI into the analysis of the MINOS experiment by analytic arguments and by using the GLoBES software in order to simulate how the sensitivity to the ordinary neutrino oscillation para\-meters is affected by the introduction of NSI. We have also studied the prospects of putting bounds on the effective NSI parameters directly from the MINOS data.

Our analytic results show that the disappearance channel ($\nu_\mu \rightarrow \nu_\mu$ oscillations) is mainly affected by the effective NSI parameter $\eps_{\tau\tau}$, while the appearance channel ($\nu_\mu\rightarrow \nu_e$) is mainly sensitive to $\eps_{e\tau}$. The effect of including NSI into the analysis of the disappearance channel is that the sensitivity contours are extended to larger $\Delta m_{31}^2$ and lower $\sin^2(2\theta_{23})$ by the introduction of a degeneracy due to $\eps_{\tau\tau}$ as described in the analytic treatment. These analytic considerations are supported by our numerical simulations, which are performed using the full three-flavor framework.

In the numeric analysis of the appearance channel, the degeneracy between the leptonic mixing angle $\theta_{13}$ and the effective NSI parameter $\eps_{e\tau}$ described in \Refs~\cite{Huber:2002bi,Blennow:2005qj} introduces difficulties in placing bounds on either of these parameters unless a stringent bound for the other parameter is imposed by external measurements. With an external bound on $|\eps_{e\tau}|$ of the order of $10^{-1}$, the MINOS experiment would be sensitive to values of $\sstt$ down to about 0.07, to be compared with the present bound of approximately 0.13 from the CHOOZ experiment. However, this sensitivity rapidly deteriorates with less stringent bounds on $|\eps_{e\tau}|$ and the $\sstt$ sensitivity for the NSI parameters, which are phenomenologically viable today, is clearly worse than the CHOOZ bound. On the other hand, if external measurements show that $\sstt \lesssim 0.01$, then the MINOS experiment should be able to place a bound on $|\eps_{e\tau}|$ of the order of unity, which is of the same order as the present bounds from interaction experiments. With the current CHOOZ bound, the bound that could be put on $|\eps_{e\tau}|$ from the MINOS experiment is about a factor of 2.5 larger than this, even with an increased running time of 15 years. However, a signal in the MINOS appearance channel would indicate that either $\theta_{13}$, $\eps_{e\tau}$, or both are non-zero as this would imply $\tilde U_{e3} \neq 0$.

It should also be noted that the results presented here do not depend on the neutrino mass hierarchy (\ie, if $\Delta m_{31}^2 > 0$ or $\Delta m_{31}^2 < 0$), or whether or not we try to make a fit with the same mass hierarchy as the one used in the simulation.

In conclusion, it seems that the MINOS experiment is very close to being able to put a useful bound on $|\eps_{e\tau}|$ if $\sstt$ could be further constrained by, \eg, future reactor experiments. Thus, the next generation of neutrino oscillation experiments should be able to put bounds on $|\eps_{e\tau}|$ which are more stringent than the direct bounds from interaction experiments.

\begin{acknowledgments}
We wish to thank Mark Rolinec for providing the modified AEDL files and Walter Winter for providing the updated GLoBES version.

This work was supported by the Royal Swedish Academy of Sciences
(KVA) [T.O.] and the Swedish Research Council (Vetenskapsr{\aa}det), Contract
No.~621-2005-3588.
\end{acknowledgments}

\end{document}